\pdfoutput=1
\documentclass{sigchi}

\toappear{\scriptsize Permission to make digital or hard copies of all or part of this work for personal or classroom use is granted without fee provided that copies are not made or distributed for profit or commercial advantage and that copies bear this notice and the full citation on the first page. Copyrights for components of this work owned by others than the author(s) must be honored. Abstracting with credit is permitted. To copy otherwise, or republish, to post on servers or to redistribute to lists, requires prior specific permission and/or a fee. Request permissions from Permissions@acm.org\\
\\ 
{\emph{DIS '20, July 6–10, 2020, Eindhoven, Netherlands} } \\
\copyright~2020 Copyright is held by the author/owner(s). Publication rights licensed to ACM.
\\ 
ACM ISBN 978-1-4503-6974-9/20/07 
\$15.00.
\\
http://dx.doi.org/10.1145/10.1145/3357236.3395444}

\usepackage{balance}       
\usepackage{graphics}      
\usepackage[T1]{fontenc}   
\usepackage{txfonts}
\usepackage{mathptmx}
\usepackage{color}
\usepackage{booktabs}
\usepackage{textcomp}
\usepackage{comment}

\usepackage[utf8]{inputenc}
\usepackage[T1]{fontenc}

\newcommand{\squeeze}[1]{\textls[-10]{#1}}
\newcommand{\squeezemore}[1]{\textls[-20]{#1}}

\usepackage{xspace}
\usepackage{microtype}        
\usepackage{ccicons}          

\usepackage{todonotes}

\usepackage{hyperref}

\def\plaintitle{From Ancient Contemplative Practice to the App Store: Designing a Digital Container for Mindfulness}

\def\emptyauthor{}
\def\plainkeywords{Mindfulness; Spirituality; Positive psychology; Mental health;}

\usepackage[font={bf,small},skip=2pt plus 3pt,position=below]{caption}
\makeatletter
\def\url@leostyle{%
  \@ifundefined{selectfont}{
    \def\UrlFont{\sf}
  }{
    \def\UrlFont{\small\bf\ttfamily}
  }}
\makeatother
\def\pprw{8.5in}
\def\pprh{11in}

\definecolor{linkColor}{RGB}{6,125,233}
\hypersetup{%
  pdftitle={\plaintitle},
  pdfauthor={\emptyauthor},
  pdfkeywords={\plainkeywords},
  pdfdisplaydoctitle=true, 
  bookmarksnumbered,
  pdfstartview={FitH},
  colorlinks,
  citecolor=black,
  filecolor=black,
  linkcolor=black,
  urlcolor=linkColor,
  breaklinks=true,
  hypertexnames=false
}

\begin{document}

\title{\plaintitle}

\numberofauthors{1}
\author{%
  \alignauthor{Kai Lukoff\textsuperscript{1}\textsuperscript{2}, Ulrik Lyngs\textsuperscript{3}, Stefania Gueorguieva\textsuperscript{2}, Erika S. Dillman\textsuperscript{2}, Alexis Hiniker\textsuperscript{2}, \linebreak Sean A. Munson\textsuperscript{2}\\
    \affaddr{\textsuperscript{1}Corresponding author: kai1@uw.edu}\\
    \vspace{0.5mm}
    \affaddr{\textsuperscript{2}DUB Group, University of Washington, US}
    \\
    \vspace{0.5mm}
    \affaddr{\textsuperscript{3}Department of Computer Science, University of Oxford, UK}
    }
}

\maketitle


\begin{abstract}
Hundreds of popular mobile apps today market their ties to mindfulness. What activities do these apps support and what benefits do they claim? \squeeze{How do mindfulness teachers, as domain experts, view these apps? We first conduct an exploratory review of 370 mindfulness-related apps on Google Play, finding that mindfulness is presented primarily as a tool for relaxation and stress reduction. We then interviewed 15 U.S. mindfulness teachers from the therapeutic, Buddhist, and Yogic traditions about their perspectives on these apps. Teachers expressed concern that apps that introduce mindfulness only as a tool for relaxation neglect its full potential. We draw upon the experiences of these teachers to suggest design implications for linking mindfulness with further contemplative practices like the cultivation of compassion. Our findings speak to the importance of coherence in design: that the metaphors and mechanisms of a technology align with the underlying principles it follows.}
\end{abstract}

\ccsdesc[500]{Human-centered computing~Empirical studies in HCI}

\keywords{\plainkeywords}

\squeeze{\printccsdesc}

\section{Introduction}
Mindfulness, often defined as non-judgmental awareness of the present moment \cite{Kabat-Zinn2013-wm}, can improve lives. In therapeutic practice, mindfulness has been found to promote physical and mental well-being \cite{Creswell2017-px}. In Buddhist tradition, mindfulness is often seen as the foundation for ethical and wisdom trainings that develop compassion and introspection \cite{Dahl2018-fc,Dahl2015-am}. 

\squeeze{Digital technology has great potential to bring mindfulness to wider audiences for whom the practice would otherwise be inaccessible \cite{Davidson2018-rh}. However, designing digital technologies for mindfulness is also a grand challenge \cite{Shneiderman2016-by}. Adapting \textit{any} offline intervention to an online tool requires difficult decisions about questions such as how to adapt the social experience and safety features in the absence of face-to-face interaction \cite{Pagoto2016-dd}. Moreover, although it is often treated as a modern health therapy, mindfulness is also an ancient contemplative practice with over 2,000 years of history, which raises particular questions about how faithful mindfulness technologies ought to be to the original tradition. Are designers leveraging popular mobile UI design patterns in mindfulness apps--such as ads, streaks, and badges? And if so, do these fit with the teachings of mindfulness?}

This paper examines how design practitioners have taken up the challenge of creating mobile apps that teach or facilitate mindfulness and how mindfulness teachers view the experiences these apps facilitate. Mobile apps that market their ties to mindfulness have become popular on app stores, reaching perhaps the widest audience of any digital technology for mindfulness. Apps like \textit{\href{https://play.google.com/store/apps/details?id=com.getsomeheadspace.android&hl=en_US}{Headspace}} and \textit{\href{https://play.google.com/store/apps/details?id=com.calm.android&hl=en_US}{Calm}} claim to teach mindfulness to tens of millions of users. Others like \textit{\href{https://play.google.com/store/apps/details?id=com.selfhealing.candle11}{Night Candle}} and \textit{\href{https://play.google.com/store/apps/details?id=com.appcraft.unicorn}{UNICORN}} aim to facilitate mindfulness through activities such as listening to relaxing sounds and coloring. 

We first conducted an exploratory app review, in which we coded 370 mobile apps with the term ``mindfulness'' in their title or description to answer two research questions:
\begin{itemize}
\setlength\itemsep{-0.5em}
\item \textit{RQ1.1: What  user  activities  are  present in mindfulness-related mobile apps?}
\item \textit{RQ1.2: What outcomes are claimed by mindfulness-related mobile apps?}
\end{itemize}
Although prior work has reviewed apps for meditation, here we examine the wide range of apps that claim to support mindfulness in all forms and activities. Our review identifies user activities that sometimes explore the boundaries of what has traditionally been considered a mindfulness practice. In particular, we find that mindfulness is commonly presented as a tool for sleep and relaxation.

We then interviewed 15 mindfulness teachers in the U.S. to elicit expert evaluations of what kinds of experiences apps with these divergent features are likely or unlikely to offer relative to traditional mindfulness practice. This addressed the following research questions:

\begin{itemize}
\itemsep-0.1em
    \item \textit{RQ2.1:\ What do mindfulness teachers think about existing mindfulness-related mobile apps?}\\
    Because many apps in our review supported relaxation activities, we were particularly keen to understand how mindfulness teachers viewed the relationship between mindfulness and relaxation. We also asked teachers to review common features in mindfulness-related apps like streaks and badges to see where they considered them to be congruent or incongruent with mindfulness practice.
    
    \item \textit{RQ2.2:\  What might technology designers learn from  mindfulness teachers' teaching practices?}\\
    In particular, we wanted to understand how teachers introduce mindfulness not just as a self-regulation technique for stress reduction, but also as the foundation for further contemplative practices. What opportunities does this suggest for mindfulness technologies that also seek to encourage qualities like kindness or self-insight?
\end{itemize}

Our investigation suggests design implications for how designers can adapt mindfulness to the affordances of mobile apps while still retaining the benefits of traditional means of practice that motivate its adoption in the first place. We identify underexplored opportunities for technologies to link mindfulness to further contemplative practices like the cultivation of compassion and introspection, one of its powerful roles in the Buddhist tradition. Finally, we call attention to a need for coherence in design: that the underlying principles of a technology and its features make sense together.

\section{Related Work}
\enlargethispage{3\baselineskip}
To point the reader to what is changing as mindfulness moves into mobile apps, we first introduce three different conceptualizations of mindfulness and a debate over whether mindfulness can be detached from its traditional ties to other contemplative practices. We then review how mindfulness has been addressed in HCI research and commercial mobile apps.

\subsection{Three Conceptualizations of Mindfulness}
There is no consensus definition of mindfulness in Buddhism, therapy, or psychology. Instead, mindfulness is an umbrella term for a variety of practices and processes that are \textit{``largely defined in relation to the capacities of attention, awareness, memory/retention, and acceptance/discernment''} \cite[p.2]{Van_Dam2017-nn}. In a recent meta-review \cite{Terzimehic2019-eo}, Terzimehić et al. similarly found a variety of mindfulness definitions in HCI research. They characterize three mindfulness conceptualizations relevant to HCI, which we draw upon here.

\textbf{In the Buddhist tradition}, mindfulness is a translation of the Pali word \textit{sati}, which can also mean ``to recollect'' or ``to remember'' \cite{Sharf2015-xs}. In this conceptualization, mindfulness entails awareness of the present moment while at the same time recalling the Dharma, i.e., the Buddha’s ethical and spiritual teachings \cite{Cox1992-iq,Gethin2001-xs}. In the early 20th century, the Buddhist modernism movement opened the door for ordinary people to practice mindfulness, a radical departure from earlier Buddhist lineages that required scholarship of Buddhist texts and a monastic life \cite{Sharf2015-xs}. Today, Buddhist modernism mindfulness does not require people to become monks, but is still part of a web of ethical and spiritual considerations \cite{Dahl2018-fc}.

\squeeze{\textbf{In therapeutic practice}, mindfulness is taught as a tool that helps people self-regulate their attention and thereby manage health issues like stress and anxiety \cite{Vago2012-pq}. In the late 1970s, Jon Kabat-Zinn pioneered this secularized, therapeutic approach by developing the 8-week Mindfulness-Based Stress Reduction (MBSR) program, which includes exercises based on Theravada Buddhism meditation practices and Hatha Yoga body movements \cite{Praissman2008-ka}. Kabat-Zinn authored one definition of mindfulness that is common in popular media, scientific research, and HCI \cite{Terzimehic2019-eo}: \textit{``Mindfulness means paying attention in a particular way: on purpose, in the present moment, and nonjudgmentally''}} \cite{Kabat-Zinn2013-wm}. 

\textbf{In psychological research}, one attempt at a consensus definition is Bishop et al.'s two-component model \cite{Bishop2006-gt} in which mindfulness is a state that entails: 1) \textit{the self-regulation of attention} towards experience in the present moment; and 2) \textit{an orientation} towards that experience characterized by curiosity, openness, and acceptance. However, other conceptualizations are also popular in psychology research, such as mindfulness as a state of creativity \cite{Pirson2012-zl}, noting or describing \cite{Baer2004-iw,Baer2006-me}, or non-reactivity to distressing thoughts \cite{Baer2006-me}. These definitions share a focus on mindfulness as a psychological construct, whereas the therapeutic conceptualization emphasizes the role of mindfulness in the promotion of health.

Throughout this paper, we use these three conceptualizations to orient the reader to different points in the design space for mindfulness technologies.

\subsection{Mindfulness as a Precursor to Contemplative Practices}
In therapy and psychological research, mindfulness is often viewed as a standalone practice or construct. By contrast, in the Buddhist conceptualization, mindfulness is part of a bundle of ethical and wisdom trainings \cite{Horn2015-mm}, which can collectively be referred to as contemplative practices.

Dahl and Davidson define contemplative practices as \textit{``efforts that promote human flourishing by training the mind''} \cite[p.60]{Dahl2018-fc}, and group them into three schools:

\begin{enumerate}
\itemsep-0.1em
	\item \textit{Attentional practices} train focus and awareness. E.g., mindfulness meditation, mantra recitation.

	\item \squeezemore{\textit{Constructive practices} cultivate prosocial and virtuous qualities. E.g., compassion meditation, loving-kindness meditation.}

	\item \squeezemore{\textit{Deconstructive practices} elicit self-inquiry and insight. E.g., analytical meditation, Zen Koans (riddles that provoke doubt).}
\end{enumerate}

In contemplative practices, attentional practices have traditionally been viewed as a precursor for constructive and deconstructive practices \cite{Dahl2018-fc}. For example, Buddhist practice often starts with mindfulness meditation which trains students to recognize presence of emotions like anger in the mind (i.e., noticing ‘I feel angry’). Deconstructive meditations may then leverage the ability to sustain attention to investigate the components of this anger, question one's attachment to it, and even inquire into what it means about the nature of the self \cite{Dahl2015-am}. In this view, mindfulness is not merely a therapeutic technique to relieve stress but a practical method for supporting human well-being and self-transformation \cite{Walsh2016-du, Akama2015-ey}, that complements and overlaps with positive psychology \cite{Ivtzan2016-cq}.

There is lively debate over whether mindfulness ought to be ``unbundled'' from the rest of contemplative practices in Buddhism, as it is in therapy \cite{Horn2015-mm}. Some credit this move for helping mindfulness reach a wider audience while others critique how the singular focus on mindfulness neglects other contemplative practices \cite{Walsh2016-du}. This unbundling is evident not only in therapeutic practice, but also in many of the mindfulness technologies that we turn to next.


\subsection{Mindfulness in HCI Research}
\enlargethispage{1\baselineskip}
According to Terzimehić et al.'s meta-review \cite{Terzimehic2019-eo}, mindfulness has been ascribed three different roles in HCI research: mindfulness as a mediator of other outcomes, a goal at the end of the road, or a way of being. Here, we first review HCI research that aligns with each of these three roles before revisiting how each might be supported, challenged, or informed by the way mindfulness teachers think and teach.

\squeeze{First, mindfulness is often seen as a \textbf{means} to reach another outcome. This perspective follows from therapeutic practice, where mindfulness programs are often explicitly designed to achieve a specific outcome (e.g., the MBSR program for stress reduction). HCI research has targeted outcomes such as stress using soundscapes that respond to brainwaves \cite{Cochrane2019-rf,Cochrane2018-yv} and breathing patterns \cite{Vidyarthi2012-ks,Vidyarthi2013-no}, breathing rate via haptic and audio feedback \cite{Paredes2018-dv}, and chronic pain with virtual reality systems \cite{Gromala2015-td}. Kavous et al. propose a framework for how interactive technologies like these can provide audio, visual, and touch feedback: mindfulness meditation triggers a \textit{relaxation response}, which technology can detect and respond to with \textit{restorative} stimuli that avoid tired patterns or stimulating judgment \cite{Salehzadeh_Niksirat2017-mj}.} 

\squeeze{Second, mindfulness is sometimes considered a \textbf{goal} at the end of the road. In this view, a student can practice reaching a short-term state of mindfulness, and in doing so build their capacity to be mindful in everyday life \cite{Brown2003-fa}, i.e., mindfulness as a character trait. However, mindfulness as a goal is never truly achieved–the road never ends–because mindfulness must still be practiced at each moment.} \squeezemore{Researchers have created HCI technologies that support moments of mindfulness. For instance, the \textit{Mind Pool} \cite{Long2013-mr} used ambiguous feedback and the kitchen blenders in \cite{Van_Rheden2016-fz} were redesigned to control blending speed with a pull string to promote mindful engagement. \textit{The Spheres of Wellbeing} were co-created with mental health patients to develop mindfulness within the context of Dialectical Behavioral Therapy \cite{Thieme2013-bb}. These technologies do not work towards an immediate goal like stress or pain reduction, but may cultivate mindfulness as a character trait that also has beneficial correlates.}

\squeeze{Third, mindfulness is also viewed as a \textbf{way of being} in the world. This research perspective has explored how artifacts or environments can facilitate mindfulness, rather than teaching it explicitly. Akama and Light present a cigarette and \textit{torii} gate (the entrance to a shrine) as two artifacts that serve as personal ``portals'' to moments of mindfulness by breaking the routines of everyday life \cite{Akama2015-ey}. Zhu et al. contrast instrumental artifacts, such as cushions, candles, or guided meditations, \textit{through which} one can practice mindfulness against non-instrumental artifacts, such as flowers or paintings, \textit{in or with which} one can be mindfully present \cite{Zhu2016-rq}.} Terzimehić et al. question whether artifacts can really carry or evoke mindfulness and what this means for creating 'reproducible' mindfulness experiences \cite{Terzimehic2019-eo}.




\subsection{Mindfulness in Mobile Apps}

Technology can make mindfulness more accessible. On a practical level, mobile apps are convenient, allowing users to practice anytime and anywhere, and affordable \cite{Dahl2018-fc}. They can also offer a variety of teachers, content, and meditation styles.
 
However, some mobile apps promote conceptualizations of mindfulness that are contested. The term ``McMindfulness'' has been used to describe a ``cottage industry where profit motive appears to contradict the ethical foundation of the practice it appropriates'' \cite{Healey2013-oh}, a critique which has been directed at mobile apps that present mindfulness as a series of five-minute relaxation exercises \cite{Purser2019-ja}.

A few studies have reviewed features and usability of mindfulness apps on app stores. Plaza et al. evaluated 50 mindfulness apps \cite{Plaza2013-dv}, finding that nearly 60$\%$  were dedicated to daily meditation practice and that the most common features across all apps were reminders, tracking statistics, audio tracks, and instructional texts. Mani et al. reviewed more than 500 mindfulness-based apps \cite{Mani2015-kv}, but only 23 met their inclusion criteria for providing mindfulness education and training, and most of those had poor usability. Roquet and Sas \cite{Dauden_Roquet2018-hj} analyzed 16 of the most popular iPhone meditation apps using a taxonomy of meditation methods \cite{Nash2013-ek}. 13 of those 16 apps taught meditation via extrinsic processes (e.g., guided audio instructions) and called for ``undesigning'' \cite{Pierce2012-wq} mindfulness apps to fade into the background (e.g., by using a simple timer) to allow more space for intrinsic, i.e. self-reliant, processes. 

These papers shed light on the features and usability of reviewed apps. However, because they review a relatively small number of apps, we do not know the extent of the diverse mindfulness conceptualizations available on app stores. For instance, by using the inclusion criteria that apps must have an educational curriculum for mindfulness meditation, Mani et al. exclude 95$\%$ of their initial search results from their detailed analysis. From an HCI perspective, the same inclusion criteria would exclude the vast majority of recent mindfulness systems in HCI research (cf. \cite{Terzimehic2019-eo})\ and thus also miss relevant emergent trends in the commercial market.

\section{Exploratory Review of 370 Mindfulness-related Mobile Apps}
\squeeze{In this review, we characterize the diversity of mindfulness-related mobile apps on the app store with respect to the user activities they support and the outcomes they claim. We discuss how mindfulness-related mobile apps push the boundaries of the Buddhist and therapeutic conceptualizations of mindfulness.}

\subsection{Methods}
\enlargethispage{1\baselineskip}
\squeeze{We selected 370 mindfulness-related mobile apps and inductively coded them for user activity and claimed outcomes.}

\squeeze{\textit{Selection criteria}. We reviewed apps for Android smartphones on the Google Play store in the United States. We chose Google Play because Android is the most popular mobile operating system in the U.S. \cite{noauthor_2020-xc} and previous reviews have found that it grants more permissions to developers than the Apple App Store, which enables developers to explore a wider variety of user interactions \cite{Lyngs2019-mw} that might reveal new possibilities for mindfulness in the digital realm. Google Play also provides information about install counts--which the Apple App Store does not--that we used to select apps that met a minimum threshold of popularity. However, there are likely some apps that are only available on other app stores (e.g., the Apple App Store) that were not included, so our review is not fully comprehensive.}

\squeeze{In March 2019, we searched for all smartphone apps that mentioned the term ``mindfulness'' in either the title or description, using the 42Matters API.\footnote{https://42matters.com/app-market-data} We chose \textit{not} to influence our search results by requiring apps to match additional search terms such as ``meditation.'' Our search thus cast a wide net to capture the diversity of mindfulness conceptualizations. We refer to apps in our corpus as ``mindfulness-related apps'' to clarify that mindfulness was not necessarily the primary activity of apps.}

Our search yielded 1342 apps. This was reduced to 379 apps after setting an inclusion criteria of a minimum of 5K-10K installs on Google Play (install counts are listed in ranges). We did this for two reasons. First, apps with higher installs should better reflect the experiences people have with mindfulness-related apps. Second, this sample seemed sufficient to characterize the diversity of apps, but not so large as to overwhelm our research team. Of these, 9 were either not available at the time of coding or unavailable in English, leaving 370 apps.

\textit{Coding process}. Following similar reviews, we coded apps for the user activities and outcomes explicitly promoted in descriptions, screenshots, and videos on the Google Play store \cite{Shen2015-vd, Stawarz2015-zn, Stawarz2018-mi}. User activities included those that the app provided on the phone (e.g., reading inspirational quotes) or supported off the phone (e.g., seated meditation with guided audio). Outcomes included any effects that the app promoted in its materials (e.g., improved attention).

The first author and three colleagues reviewed a random selection of 50 apps to independently and inductively identify potential codes \cite{Elo2008-hb}. These researchers then met to discuss and develop a single set of codes. Each researcher then independently coded 80 additional apps, resolving any ambiguous cases by discussing the app with the other members of the coding team. Our final codebook is available here:
\href{https://osf.io/b9jwu}{https://osf.io/b9jwu}

\textit{Deep dive reviews}. In parallel with the coding process, our research team selected 16 mindfulness-related apps with diverse user activities (e.g., sitting meditation, sleeping, and coloring) to experience in-depth.\footnote{For a list of all 16 deep dive review apps, see: \href{https://osf.io/b9jwu}{https://osf.io/b9jwu}} We installed each app, used it at least three times, and then discussed it at our weekly group meetings. We paid particular attention to UI patterns that we saw as potentially diverging from the Buddhist and/or therapeutic conceptualizations of mindfulness. 

\subsection{Results}
\enlargethispage{1\baselineskip}
\squeezemore{We characterize the 370 most-installed mindfulness-related mobile apps on Google Play. The install count of apps ranged from 5K-10K to 10M-50M (median: 10K-50K). Half were in the Google Play category of Health and Fitness (50$\%$), with Lifestyle (11$\%$), Entertainment (6$\%$), Education (6$\%$), Puzzle (5$\%$), and a long tail of other categories (22$\%$) comprising the remainder.}

\subsubsection{User Activities}
\begin{figure}
    \centering
    \includegraphics[width=\linewidth]{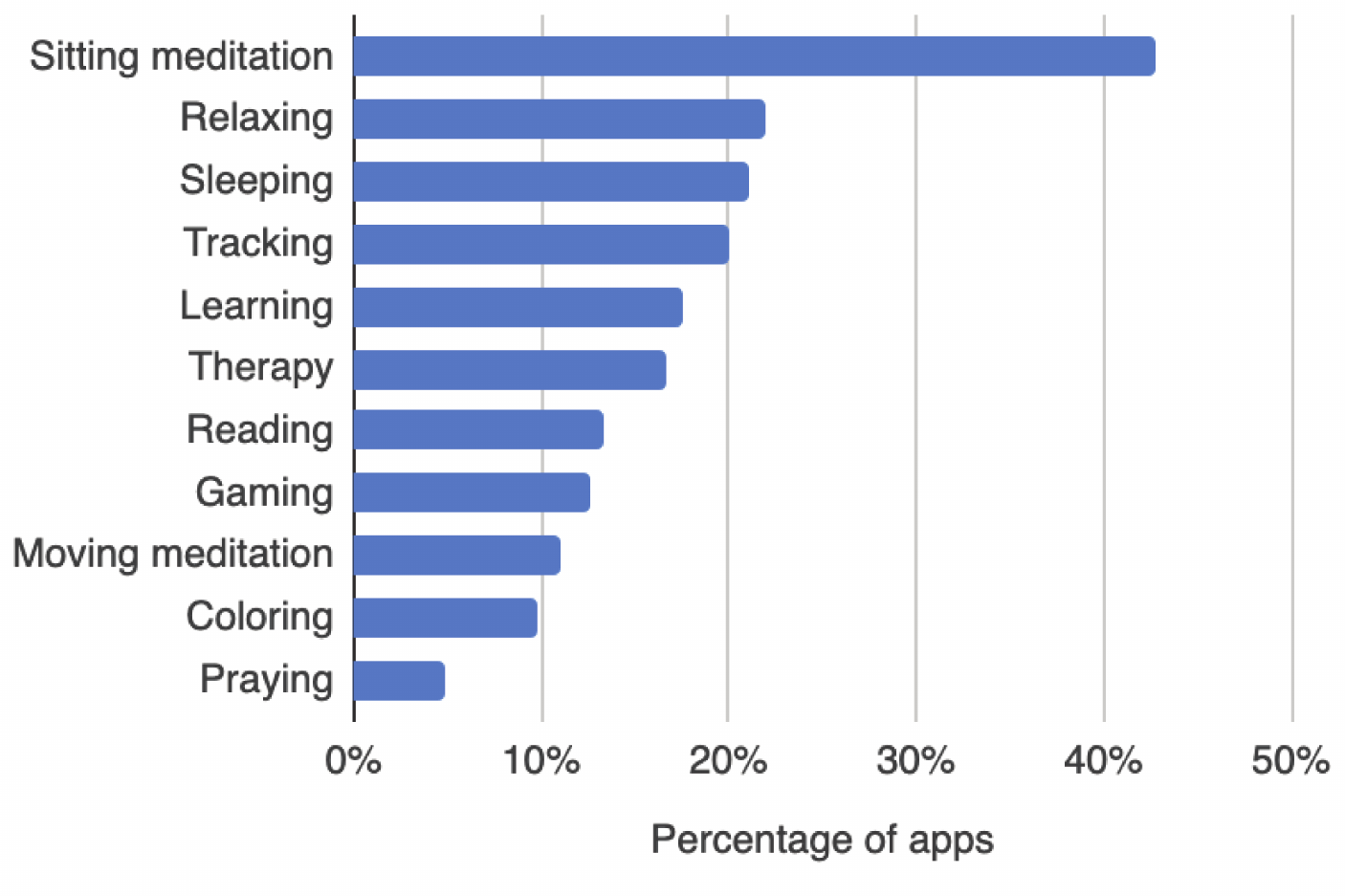}
    \setlength{\abovecaptionskip}{-10pt} 
    \setlength{\belowcaptionskip}{-10pt} 
    \caption{User activities supported by mindfulness-related apps (n=370)}
\label{fig:user_activities}
\end{figure}

\textbf{Figure \ref{fig:user_activities}} shows the percentage of apps that supported each user activity. One app could be coded for multiple activities.\footnote{For a description of each user activity, key features that support that activity, and example apps, see: \href{https://osf.io/b9jwu}{https://osf.io/b9jwu}} Sitting meditation (43$\%$) was by far the most frequent user activity, with the key features of these apps --- audio-guidance, video-guidance, and timers --- mirroring findings in previous reviews \cite{Sliwinski2017-ac,Zhu2017-fm}. Relaxation (22$\%$) and sleep (21$\%$) were the next most common. Activities that are not traditional mindfulness practices included relaxing, sleeping, tracking (21$\%$), and gaming (13$\%$). 

\subsubsection{Claimed Outcomes}
\begin{figure}
    \centering
    \includegraphics[width=\linewidth]{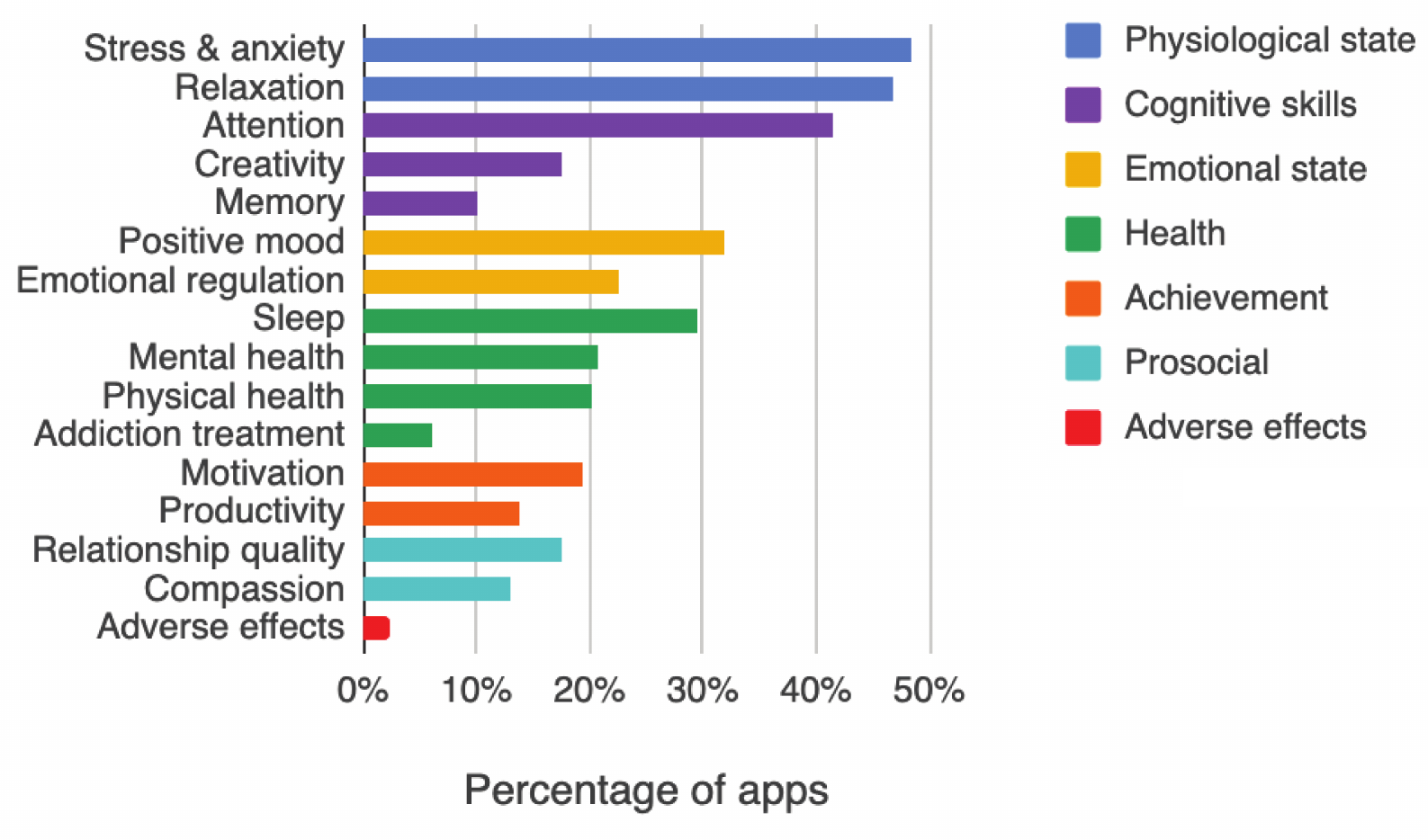}
    \setlength{\abovecaptionskip}{-10pt} 
    \setlength{\belowcaptionskip}{-20pt} 
    \caption{Claimed outcomes of mindfulness-related apps (n=370)}
\label{fig:claimed_outcomes}
\end{figure}

\textbf{Figure \ref{fig:claimed_outcomes}} shows the outcomes claimed in the app descriptions. One app could be coded for multiple outcomes. We clustered the outcomes into the following higher-level groupings:

\enlargethispage{1\baselineskip}

\begin{itemize}
\itemsep-0.1em
    \item \textbf{Physiological state:} Stress and anxiety reduction (48$\%$ of apps) and relaxation (47$\%$) were the two most frequently claimed benefits.

    \item \textbf{Cognitive skills: }Sharpened attention (41$\%$) as well as claims to creativity (18$\%$) and memory (10$\%$).

	\item \textbf{Emotional state:} Positive mood (32$\%$) and skills for emotional self-regulation (23$\%$).

	\item \textbf{Health:} Sleep (29$\%$), mental health (21$\%$), physical health (20$\%$), and addiction treatment (6$\%$).

	\item \textbf{Achievement:} Motivation (19$\%$) and productivity (14$\%$).

	\item \squeezemore{\textbf{Prosocial:} Relationship quality (18$\%$) and compassion (14$\%$).}

	\item \textbf{Adverse effects:} \squeezemore{Adverse effects (2$\%$) included any negative effects or warnings on the app’s store page, including disclaimers that the app is not a replacement for medical help.}
\end{itemize}

\subsubsection{Deep dive reviews}
Across the 16 apps in our deep dive reviews, we identified three types of features that seemed to deviate from the Buddhist and/or therapeutic conceptualizations of mindfulness. Some tracking features, like progress bars, showed mindfulness as an achievement level rather than as a state that is sustained moment-to-moment. Similarly, gamification features like badges suggested mindfulness as a series of achievements rather than as an never-ending journey or a way of being. Finally, social features sometimes prompted users to compare themselves against others (e.g., in terms of daily meditation streaks). In Study 2, we asked mindfulness teachers for their perspective on these features.

\subsection{Analysis of results}
\textit{Varied Interpretations of Mindfulness on the App Store}\\
Similarly to the common academic use of `mindfulness' as an umbrella term, mindfulness-related apps offer a wide range of user activities (from meditation to prayer) and claimed outcomes (from stress reduction to productivity).

\squeeze{However, not all notions of mindfulness are equally represented on the Google Play store. 48$\%$ of apps claimed relaxation as an outcome and 47$\%$ did so for stress \& anxiety reduction, which aligns with the therapeutic conceptualization of mindfulness. Only about 15$\%$ promoted prosocial outcomes as a benefit, although they are a common aim of the contemplative practices linked to mindfulness in the Buddhist conceptualization.}

Relaxing and sleeping were among the most frequent user activities supported by mindfulness-related apps, although these are not traditionally part of mindfulness practice. Relaxation and improved sleep have long been recognized as correlates to mindfulness practice \cite{Baer2003-hd}, but in many mindfulness-related apps they seem to have become the primary activity. \textit{Headspace} and \textit{Calm} for instance, started as meditation apps, but now boast large collections of bedtime stories, some read by celebrities \cite{Constine2019-zj}. Around half of \textit{Calm's} users now use it primarily for sleep, rather than for meditation  \cite{Potkewitz2018-kd}.

Some apps stretch the fabric of the mindfulness umbrella. Coloring apps, for instance, hold promise as a digital version of the mindfulness practice of mandala coloring. However, many have been found to turn coloring into a series of simple taps to fill in an image, which eliminates the need for skilled attention that is at the heart of the original practice \cite{Dauden_Roquet2019-mk}. Other apps promise benefits like productivity that are not traditionally emphasized within mindfulness practice. The app \textit{\href{https://play.google.com/store/apps/details?id=com.amazingapps.millionairemindestcourse&hl=en_US}{	
Millionaire Mindset}} provides a chapter on mindfulness in a course that is \textit{``not only about how to get rich, but how to STAY rich.''} 

\textit{Unsubstantiated Benefits, Understated Risks}\\
\squeeze{The benefits claimed by mindfulness-related apps reach beyond what current evidence supports. A 2014 meta-analysis \cite{Goyal2014-ak} found that mindfulness-based interventions (relative to controls) showed moderate efficacy in treating anxiety, depression, and pain and low efficacy in reducing stress and improving quality of life. Similarly, a meta-analysis of 43 technology-enabled mindfulness-based interventions found that they decreased negative affect \cite{Victorson2020-bh}. However, neither study found evidence for other benefits that mindfulness-related apps claim such as sleep or positive mood. Moreover, these results are generally based upon large ``doses'' of mindfulness in demanding programs like MBSR, which lasts for 8 weeks and over 20 hours, and cannot be equated with a 5-minute meditation in an app \cite{Creswell2017-px, Van_Dam2017-nn}.}

Potential adverse effects, meanwhile, were stated in the materials for only 2$\%$ of all apps. Mindfulness-related apps follow a widespread assumption that mindfulness meditation is associated with minimal, if any, adverse effects. However, this conclusion is based primarily on a lack of research rather than on evidence of absence \cite{Van_Dam2017-nn}. A recent study documents incidents of meditation-related experiences that are difficult, distressing, and functionally impairing \cite{Lindahl2017-kl}, with the last author commenting, \textit{``The fact that I am receiving calls from meditators-in-distress indicates that they are not receiving adequate support from wherever they learned to meditate''} \cite{noauthor_2019-lo}. Overstated benefits and understated risks have been reported in other health-related apps \cite{Grundy2016-vy} and our results suggest this issue may be prevalent in mindfulness-related apps, too.

Designers of technologies for mindfulness mediation should consider how to address potential adverse effects. Sitting meditation was supported by 43$\%$ of apps in our sample, but only 2$\%$ of those contained a mention of adverse effects. At a minimum, a disclaimer in promotional materials and/or during on-boarding might be warranted. 
However, we encourage designers to go further and adapt best practices from the Meditation Safety Toolbox \cite{Britton2019-ce}. For instance, for students affected by trauma, guidelines advise against open awareness meditations, where students turn attention to anything that arises in consciousness without reacting to it, which may overwhelm them with troubling thoughts. A focused attention technique, where students focus on a specific stimulus like the breath, would be more appropriate. However, none of the meditation apps that we reviewed in our deep dive reviews screened for such health concerns.

\section{Interviews with Mindfulness Teachers}
Our interviews provided an understanding of mindfulness teachers' expert perspectives on existing mindfulness apps, helping identify potential opportunities for designers.

\subsection{Methods}
\enlargethispage{1\baselineskip}
\textit{Recruitment}. We recruited in Seattle, Washington, US, by emailing and calling organizations that self-describe as teaching mindfulness and sharing our recruitment message via social media and email lists. In a screening survey, we asked respondents about their years of teaching experience, teaching tradition, training received, where they teach, and whether they currently used any mindfulness-related apps on their phones. The interviews were approved by the University of Washington Human Subjects Division.

\squeezemore{We characterize teachers as belonging to one of three traditions}:
\begin{itemize}
\itemsep-0.1em
	\item \textbf{Therapeutic:} Teach the therapeutic conceptualization of mindfulness; included teachers who taught MBSR at a university and a medical clinic and another who taught mindfulness for chronic pain management at a hospital.

	\item \textbf{Buddhist:} Teach the Buddhist conceptualization of mindfulness; included teachers who taught in the Zen, Theravadan, and Insight Meditation traditions.

	\item \textbf{Yogic:} \squeeze{Embeds mindfulness in yoga--a form of movement and body awareness--that is a major component of the MBSR program \cite{Praissman2008-ka}; included teachers who taught various types of yoga (e.g., Hatha, Vinyasa, Nidra). All of these teachers listed their yoga as a form of mindfulness practice and also taught meditation integrated into their yoga and/or separately.}
\end{itemize}

This designation of teaching tradition helped us understand the differences in perspectives between teachers and we report it when we quote participants (e.g., P6\textsubscript{T} for therapeutic). However, there was often crossover in a teacher’s training and teachings--for instance, three teachers had trained extensively in the Buddhist tradition but were now teaching with a therapeutic orientation. 

We refer to participants collectively as ``mindfulness teachers,'' although some identify primarily as a chaplain, physical therapist, or yoga teacher. We refer to the people they teach as ``students,'' although some identify as clients or patients. We clarify these roles at times when this context helps in the interpretation of results and is unlikely to compromise the identity of a participant.

\textit{Interview method and analysis}. 
\squeezemore{We conducted individual interviews using a semi-structured protocol. Interviews addressed the teacher's conceptualization of mindfulness, which mindfulness practices they teach, and potential benefits and risks of technology in mindfulness practice. We also showed teachers eight paper mockups of mobile features for tracking, gamifying, or socializing that we found in our earlier deep dive reviews of 16 apps. As these often diverged from our own understanding of the Buddhist and/or therapeutic conceptualizations of mindfulness, they served as provocations to elicit what teachers found helpful or unhelpful in mindfulness apps. Finally, if teachers used mindfulness apps themselves, we asked them to show us what they liked or disliked in each app on their phone. The interview protocol and mockups are in the supplementary materials: \href{https://osf.io/b9jwu}{https://osf.io/b9jwu}}


Interviews were held at the University of Washington or a location of interviewee’s choice (e.g., hospital). Each participant received a $\$$25 voucher. Each interview was audio recorded, transcribed, and then coded in \textit{Dedoose} software. We used a six-step process of reflexive thematic analysis \cite{Braun2006-ma, Braun2019-af, Braun2018-th}.

\subsection{Results}
We interviewed 15 teachers, 7 of whom taught in the therapeutic tradition, 4 in the Buddhist, and 4 in the Yogic. To protect anonymity, we report participant demographics in aggregate rather than individually. Interviews lasted 55-90 minutes (median: 70). Teaching experience ranged from 3-30 years (median: 9). Teachers taught at wellness programs in universities (4 teachers), schools (3), and corporations (2), at spiritual communities (3), at hospitals and health clinics (3), at yoga studios (2), and other locations.
12 teachers reported using mindfulness-related mobile apps; based on our app review, these apps included 3 apps that we coded with a primary activity of sitting meditation: Insight Timer (10 teachers), \textit{\href{https://play.google.com/store/apps/details?id=org.plumvillageapp}{Plum Village}} (1), \textit{\href{https://play.google.com/store/apps/details?id=org.wakingup.android&hl=en_US}{Waking Up}} (1); 3 apps for therapy: \textit{\href{https://play.google.com/store/apps/details?id=com.claritasmindsciences.UnwindingAnxiety&hl=en_US}{Unwinding Anxiety}} (1),
\textit{\href{https://play.google.com/store/apps/details?id=com.claritasmindsciences.EatRightNowCommercial&hl=en_US}{Eat Right Now}} (1), \textit{\href{https://play.google.com/store/apps/details?id=gov.wa.dva.maximpact&hl=en_US}{Max Impact TBI}} (1); 2 apps for moving meditation: \textit{\href{https://play.google.com/store/apps/details?id=com.glo.mobile}{Glo}} (2), \textit{\href{https://play.google.com/store/apps/details?id=com.cody.cody}{Cody}} (1); and 1 app for learning about Buddhist ethics: \textit{\href{https://play.google.com/store/apps/details?id=org.dharmaseed.android}{Dharma Seed}} (2).

\subsubsection{Thematic Analysis}
\enlargethispage{1\baselineskip}
We derived three themes from our interviews with mindfulness teachers: (1) designing the container for mindfulness, (2) the relationship between mindfulness and relaxation, and (3) mindfulness outside of formal practice.



\subsection{Designing the Container for Mindfulness}
Practicing mindfulness is hard for beginners. It can be an intimidating experience that it is less tangible than physical exercise and often unfamiliar. Students are trying to build a \textit{``mindfulness muscle''} that they have not used before (P3\textsubscript{Y}). 
To overcome this challenge, teachers often spoke of the need to create a container, a learning environment that supports formal mindfulness practice. ``Formal practice'' refers to exercises where mindfulness is the sole focus, such as meditation. By contrast, ``informal practice'' refers to cultivating mindfulness within the flow of everyday life, such as when washing dishes.



The container encompasses aspects of the formal practice environment that are both physical (e.g., floor layout, sights, and sounds) and non-physical (e.g., tone and norms). A quote by P14\textsubscript{T} captures many of the attributes of a well-designed container: \textit{``As a teacher, you set the container for the practice$ \ldots $  for me, that means keeping it safe, and {limiting or minimizing external distractions}$ \ldots $  and setting the tone or expectation for your students, that whatever they say to you is held in confidence and that they are able to express themselves without any judgment, without fear.''}

Although the container is usually referenced with respect to the environment for offline practice, the teachers in our interviews felt that it applied to the digital environment too: \textit{``If you came up with an app or a technology, that technology would have to help the person set up that container. They could set it up by having their own space or going to a quiet space to practice and things like that''} (P11\textsubscript{B}). 


\subsubsection{Characterizing an Ideal Container}
\enlargethispage{1\baselineskip}
Our analysis identified two overarching qualities of an ideal container, although the details of the design varied from teacher to teacher. First, the container should be \textbf{welcoming}, a space that invites students in. We further broke this down into four attributes:

\begin{itemize}
\itemsep-0.1em
    \item \textbf{Safe:} Create a space where people feel safe. Teachers emphasized that psychological safety was critical both in its own right \textit{and} because it opened up possibilities for students to express themselves without judgment or fear. One teacher (P15\textsubscript{T}) practiced reflective listening with students who were too distressed to practice meditation. Another (P14\textsubscript{Y}) messaged students the day after troubling thoughts came up for them in class. Several teachers expressed concern that a student using an app might not receive adequate support when such difficult experiences arise.

    \item \textbf{Accessible:} Accommodate the needs of all students. One teacher (P13\textsubscript{T}) taught students without support blocks for yoga how to use books and cans instead. Another (P8\textsubscript{B}) suggested an alternative posture for a student with scoliosis so that they would not be overwhelmed by pain during sitting meditation.

    \item \textbf{Relatable:} Make it easy for students to relate to the practice. When working with patients with Christian religious beliefs who felt uncomfortable with the word ``mindfulness,'' one teacher (P15\textsubscript{T}) instead substituted the term ``skillful awareness'' Another (P12\textsubscript{T}) brought the game \textit{Fortnite} into the classroom for teens to practice awareness during everyday activities. 

    \item \textbf{Relaxing:} Put students at ease so they can sink into the practice. This is especially important for beginners, for whom mindfulness practices like observing one's breath can be strange and trigger anxiety.
    \end{itemize}
    

Second, teachers' responses indicated that a container should be \textbf{coherent}, creating an experience of mindfulness practice that makes sense as a unified whole. Three examples illustrate how the design of the container can create a coherent--or incoherent--experience of mindfulness practice: 

\begin{itemize}
\itemsep-0.1em
    \item \textbf{Focusing, not distracting:} \squeezemore{Focusing attention is at the core of mindfulness, so it makes sense that teachers emphasized minimizing distractions in the container. Meditation retreats, for instance, are carefully designed to remove external stimulation (e.g., entertainment, social interaction) to let people observe the mind. One teacher (P14\textsubscript{Y}) found \textit{Insight Timer} to be unnecessarily busy with people virtually ``high-fiving'' each other for meditating. Another (P1\textsubscript{Y}) critiqued mindfulness apps that ``bombarded'' the user with notifications or ads.}
   
    \item \textbf{Social support, not social comparison:} In mindfulness, teachers often talk about the dangers of the ``comparing mind,'' which compels people to judge themselves against others, often self-critically \cite{Siegel2014-fd}. P11\textsubscript{B} described, \textit{``When the mind compares, it can't be present. It can't. If I'm thinking that [another student] has 37 straight days and I'm down here on the three days, then I'm not going to pay attention to my [meditation] sit.''} Teachers felt that comparing activity might work in a fitness app (P3\textsubscript{Y}) or a language-learning app (P14\textsubscript{Y}), but that it did not fit in a mindfulness app where it contradicts the mindset the practice tries to instill.

    \item \textbf{Process-oriented, not achievement-oriented:} Mindfulness practice emphasizes the moment-to-moment experience in the present, not the achievement of a goal. For teachers, this meant that mindfulness apps should avoid design metaphors that suggest linear advancement towards a goal--such as the progress bar and levels found in the meditation app \textit{\href{https://play.google.com/store/apps/details?id=com.aurahealth}{Aura}}--in favor of those that suggest a development process (e.g., a tree that can represent the student's growth). \squeeze{At a practical level, teachers were divided about whether common gamification patterns such as streaks, points, and badges would encourage students to develop mindfulness practice habits, with some believing that they would provide a helpful initial boost of extrinsic motivation. Others worried that these patterns would reduce intrinsic motivation. At a deeper level, however, many expressed concern that gamification was antithetical to mindfulness itself, which trains students to accept the present moment rather than judge it. Finally, teachers also worried that ``\textit{patronizing gold stars}'' (P4\textsubscript{T}) could trivialize mindfulness, undermining its potential as a transformative journey.} 
\end{itemize}

These three examples show how teachers worried that even a locally effective design pattern (e.g., a gamification feature that motivates students to return to the app) might be counterproductive at a global level if it contradicts the teachings of mindfulness (e.g., a focus on achievements over process). 



\subsection{The Relationship Between Relaxation and Mindfulness}

\begin{figure}
    \centering
    \includegraphics[width=\linewidth]{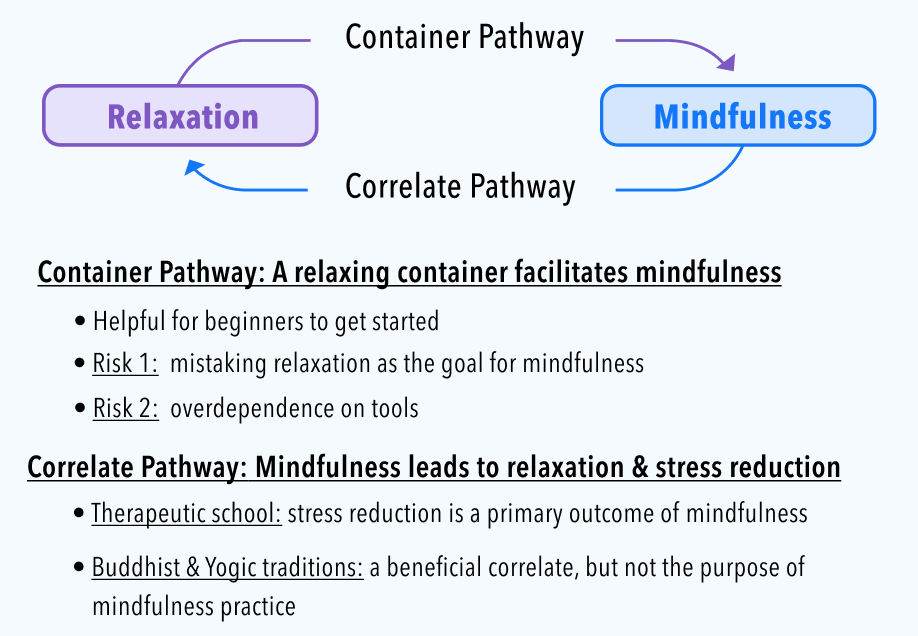}
    \setlength{\abovecaptionskip}{-10pt} 
    \setlength{\belowcaptionskip}{-20pt} 
    \caption{The relationship between relaxation \& mindfulness}
\label{fig:self_regulation}
\end{figure}

Of the mindfulness-related mobile apps we reviewed, almost half self-described as promoting relaxation. How do teachers view  the relationship between relaxation and mindfulness?

Teachers all distinguished between relaxation and mindfulness, but sometimes emphasized different relationships between the two constructs. In \textbf{Figure \ref{fig:self_regulation}}, we present a conceptual diagram of our findings from the interview data.

The \textbf{Container Pathway} illustrates how relaxation facilitates mindfulness, as a soothing environment can help beginners get started with the practice. However, teachers also cautioned about two risks. First, relaxation might be mistaken for mindfulness. Teachers did not object to relaxation per se--in fact, many teachers praised its restorative effects--but were concerned that apps might overemphasize relaxation at the expense of cultivating the moment-to-moment awareness experience at the heart of mindfulness. Moreover, without guidance, beginning students might not be able to recognize the difference: \textit{``I see this a lot in the schools that kids think that, `Oh, mindfulness means I have to be calm, or it is calm.' That's not it at all''} (P12\textsubscript{T}).

Second, the relaxation that the container provides might lead to overdependence. P14\textsubscript{Y} shared: \textit{``There can be an overreliance on always having your soothing music and always having your meditation space exactly perfect. And if you don't, then you can't meditate. Right?''} All teachers emphasized that mindfulness should extend beyond formal practice spaces to the quality of attention applied to any life activity. 


\enlargethispage{1\baselineskip}
The \textbf{Correlate Pathway} of \textbf{Figure \ref{fig:self_regulation}} shows mindfulness leading to relaxation and stress reduction, a relationship that teachers described and which is corroborated by considerable research \cite{Baer2003-hd}. In this case, however, there was a notable difference between the teaching traditions. Teachers from the therapeutic school emphasized relaxation and physiological improvements as an important outcome of their mindfulness teaching. P15\textsubscript{T} shared, \textit{``For what I can do for patients, I can introduce these practices in a very practical way that enables [patients] to feel better, to relieve suffering, to decrease their pain, decrease their anxiety, improve their symptoms.''} 


Although teachers from the Buddhist tradition, and often the Yogic, appreciated physiological improvements as a beneficial correlate of mindfulness practice, they were concerned that this focus leaves behind something important. P8\textsubscript{B}: \textit{``If your goal is to just have a sense of relaxation, that's fine. But if your goal is \ldots for it to be a transformative practice and to be able to bring it into your own daily life, then it becomes an issue.''} These teachers worried that mindfulness in the US has become so associated with self-regulation that newcomers can no longer appreciate its transformative potential. For instance, P14\textsubscript{Y} regretted that their students missed out on opportunities for self-exploration in debriefings at the end of meditation class: \textit{``And instead of questions that would come up, it was mostly, `Thank you. I feel very relaxed now.'\thinspace''} Teachers felt that mindfulness could be more than just a tool for relaxation.

\begin{figure}
    \centering
    \includegraphics[width=\linewidth]{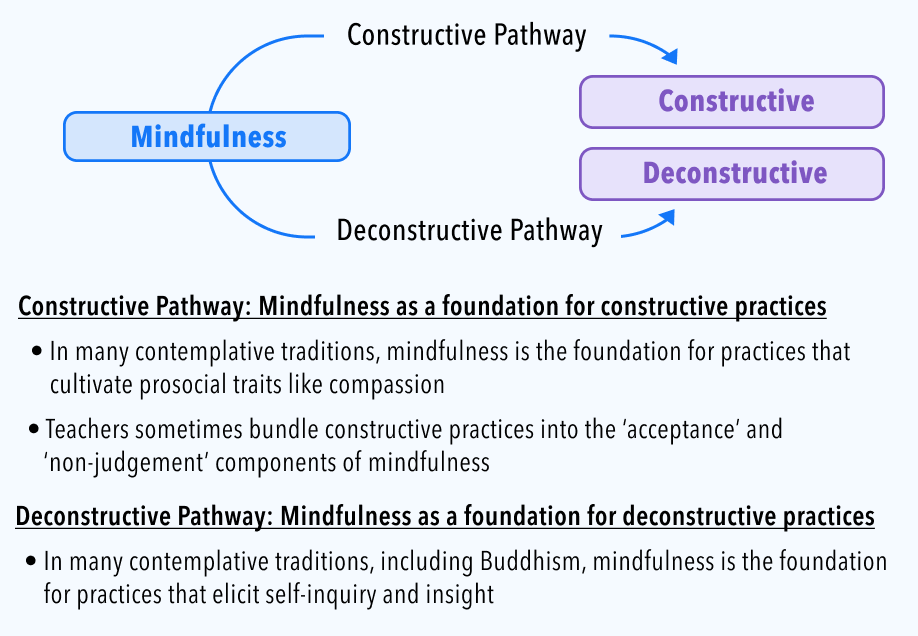}
    \setlength{\abovecaptionskip}{-10pt} 
    \setlength{\belowcaptionskip}{-20pt} 
    \caption{Mindfulness as a journey into contemplative practices}
\label{fig:journey}
\end{figure}

\squeeze{Teachers' notions of the transformative potential of mindfulness align with Dahl and Davidson's view of mindfulness as the foundation for further contemplative practices \cite{Dahl2018-fc, Dahl2015-am}. In \textbf{Figure \ref{fig:journey}}, the \textbf{Constructive Pathway} shows mindfulness facilitating constructive practices that cultivate prosocial traits such as kindness. For example, P15\textsubscript{T} used perspective-taking as a technique to develop self-compassion: \textit{``A lot of people are really hard on themselves and that just escalates their stress. I encourage people to think of how you would show up for a dear friend if they were dealing with a challenging health problem and turned to you for support. And maybe even the kind of language would you use and how it feels in your heart area.''}} 


To bridge from mindfulness to constructive practices, many teachers asked students to set intentions at the end of meditation practice. P14\textsubscript{T} described, \textit{``Sometimes [intentions] are like little stepping stones on the way to a much larger--I don't like to call them goals--but your heartfelt desire or your wish or your mission for your life, your purpose. So all these little intentions can add up and keep you on the path towards that.''} P2\textsubscript{Y} suggested that an app might scaffold prosociality by sending a just-in-time reminder for the user to practice kindness when meeting with someone who tries their patience.

The \textbf{Deconstructive Pathway} of \textbf{Figure \ref{fig:journey}} illustrates mindfulness supporting deconstructive practices for self-inquiry and insight. Teachers described ways they encouraged such introspection. P4\textsubscript{T} said, \textit{``We always ask, `What are you learning about yourself?' to the students each week.''} Other teachers introduced practices like introspective writing, journaling, and daily reflections, such as \textit{``reflecting on the limitation of being lost in thought versus being present with everything''} (P8\textsubscript{B}).

When integrated with these constructive and deconstructive practices, mindfulness looks less like a self-regulation technique for stress relief alone and more like a catalyst for the transformative journey that teachers described.

\enlargethispage{1\baselineskip}

\enlargethispage{1\baselineskip}
\subsection{Mindfulness Outside of Formal Practice}

Teachers described overreliance as a risk of of mindfulness meditation apps. As P15\textsubscript{T} describes: \textit{``The advantages of an app could be like training wheels; it's teaching somebody the basic practices and principles. I think one of the shortcomings is they never take the training wheels off, and then they become dependent: `I can only meditate if I've got my phone.'\thinspace''} Teachers expected this risk to be especially high for apps that set up a relaxing container with soothing sounds and music.

The challenge for students is to develop their own informal practice and cultivate mindfulness in everyday life, which must be done outside of the comfort of the container. Teachers prepared their students for this challenge in two ways.

First, teachers steadily decreased the amount of scaffolding within the container. Scaffolding plays an important role in mindfulness practice: relaxing music helps students overcome initial anxiety and guided instruction gently reminds students to return their focus to the breath when their minds wander. Yet there comes a time when students should no longer need any tool or teacher to practice:\textit{``Insight Timer does have guided meditations and that's wonderful. But I would want [students] to not always be listening to someone's voice guiding them through. I want my students to ultimately be able to sit with themselves and guide themselves''} (P14\textsubscript{T}). Teachers encouraged experienced students to practice without music or a timer and steadily decrease the amount of verbal instructions they provide in their own guided meditation sessions. 

Second, teachers encouraged their students to set intentions and reminders to practice outside of the container. After a formal practice like meditation, teachers often asked students to set an intention for informal practice in their daily routine (e.g., while washing the dishes (P4\textsubscript{T}) or brushing teeth (P8\textsubscript{B})). However, in keeping with the non-judgmental nature of mindfulness, they emphasized that these intentions should be softer than goals so that a student does not feel like a failure if they are unable to follow them. Intentions could also be supported by gentle reminders to practice at designated moments, which P8\textsubscript{B}\ called \textit{``meditative markers.''}

Teachers advised that both classes and apps need to be designed to help people \textit{transfer} what they learned in mindfulness practice to everyday life: \textit{``One of the important pieces is that transitional state, so something that supports that within the app. You're going through meditation, and now you've cultivated something. And it's something to now transition into the rest of your life with''} (P10\textsubscript{B}).



\section{Discussion}
Our investigation finds that mindfulness-related mobile apps support a wide range of activities that are not traditional mindfulness activities, notably sleep and relaxation. \squeezemore{Teachers felt that while mindfulness can aid with relaxation, a singular focus on relaxation and related health benefits like stress reduction may neglect mindfulness’ potential as a journey into further contemplative practices. Finally, teachers considered it important for the design patterns in apps to fit with the principles of mindfulness.}

We discuss how to (1) design for mindfulness as a journey into contemplative practice, (2) transfer mindfulness from apps to everyday life, and (3) design for coherence.



\subsection{Designing for a Journey into Contemplative Practice}
Mindfulness teachers recognized, and many praised, mindfulness technologies for their ability to train self-regulation and aid in relaxation, a topic that has also been explored in recent HCI research \cite{Cochrane2019-rf,Cochrane2018-yv,Paredes2018-dv,Salehzadeh_Niksirat2017-mj,Vidyarthi2013-no,Vidyarthi2012-ks}. A large number of studies have demonstrated the substantial health benefits of this therapeutic version of mindfulness \cite{Goyal2014-ak}, making it a valuable practice on its own, although many questions remain about how to best leverage technology as an effective delivery mechanism for such interventions \cite{Davidson2018-rh,Van_Dam2018-ep}. 

However, many teachers also felt that practicing mindfulness for this purpose alone failed to realize its full potential. Instead, they pointed to mindfulness as a transformative journey, which we connect with Dahl and Davidson's view of mindfulness as a foundation for other contemplative practices \cite{Dahl2018-fc}.

What might it look like to design technologies that integrate mindfulness with constructive and deconstructive contemplative practices? Constructive practices cultivate virtuous traits. In the Buddhist tradition, there is often a progression from attentional practices like mindfulness meditation to constructive ones like loving-kindness meditation \cite{Dahl2018-fc}, a sequencing that is found in the mobile app \textit{Headspace} and might also be employed in other technologies. 

In technologies with a more strictly secular orientation, the way to support constructive practices is less clear. One approach would be to integrate digital versions of positive psychology exercises \cite{Calvo2014-vo,Lukoff2019-kf}, such as gratitude journaling or prosocial spending. Exercises that directly leverage the attentional focus developed through mindfulness, such as vividly imagining one's ideal future self \cite{Lyubomirsky2005-vq}, might fit particularly well in such an integrated contemplative technology. Sliwinski et al. suggest digital games with a narrative as a promising way to pair mindfulness with other contemplative practices: \textit{``Guiding or enriching an experience with a meaningful narrative that includes moral dilemmas and encourages perspective taking might, moreover, benefit the development of empathy and other prosocial qualities of mindfulness''} \cite[p.1158]{Sliwinski2017-ac}.

Deconstructive practices encourage self-inquiry. Here, technology designers might draw upon the existing classroom practices of teachers and suggest that the user completes an introspective journaling exercise after completing a meditation session. Or, rather than following daily mindfulness meditation up with an inspirational quote (as in \textit{Calm}), a mobile app might present a Zen Koan, i.e., a riddle that elicits doubt \cite{Zug1967-fc}. 

\squeezemore{We hope these considerations will help designers answer a broader call within HCI to design for experiences that are transformative \cite{Kitson2019-gj} and meaningful \cite{Lukoff2018-km,Mekler2016-or}, whether grounded in spirituality/religion \cite{Bell2006-xj, Buie2013-al} or positive psychology \cite{Calvo2014-vo}.}

\subsection{Transferring Mindfulness from Apps to Everyday Life}
Teachers were concerned that students might become dependent on the scaffolding, especially in a relaxation-oriented practice, and fail to transfer their mindfulness practice into everyday life. We offer three suggestions for designers.

\squeeze{First, \textbf{design for less dependence} in the first place. Certain attributes of the container for formal practice in mobile apps might attract users and help them get started but build a counterproductive overreliance in the long run. Teachers cautioned that the relaxation features commonly found in mindfulness-related apps today might foster dependence and mislead students to mistake relaxation for mindfulness. Instead of assuming that all students need to start with maximum scaffolding, the designer of a meditation app could omit soothing sounds and visuals if beginning with a focus on the breath does not induce anxiety for the student. Instead, simple voice guidance might suffice.}

Second, \textbf{gradually reduce the scaffolding} inside of the container. In learning to practice mindfulness, teachers emphasized that sequence matters. For beginners, the music, visuals, and guided audio of mobile apps, what Zhu et. al call ``the form'' of mindfulness \cite{Zhu2017-fm}, can help create a relaxing container for the practice. Yet as students become more experienced, these same tools can become a hindrance to further practice outside of the container, where there is not always relaxing music playing. The implication is that technologies might be designed to gradually shift students from reliance upon extrinsic processes (e.g., guided audio) towards intrinsic processes (e.g., a simple timer or meditating without the app altogether).

Unfortunately, designers can often become so focused on designing for engagement that they neglect to design for disengagement from their tool \cite{tran2019modeling,Epstein2016-ts}, particularly when they face the economic pressures of growing and retaining their userbases. For the commercial designer, we highlight a tension, but regrettably cannot offer a tidy solution; mindfulness practice was not designed to be monetized. 

Third, help students \textbf{transfer mindfulness} from formal practices to informal practices and to everyday life. Terzimehić et al. note that the preponderance of HCI research to date has focused on designing for short-term rather than long-term changes in mindfulness capacity and that the latter might be fostered by encouraging mindfulness in everyday life, not just during meditation \cite{Terzimehic2019-eo}. Intentions and reminders might help bridge the divide. Mindfulness teachers often ask their students to set intentions for informal practices at particular moments during the day (e.g., when washing the dishes), an approach that some technologies already follow (e.g., \textit{Insight Timer}) and others might yet adopt. 

Gentle notifications, or \textit{meditative markers} as P8\textsubscript{B} called them, could support these intentions. In the Buddhist tradition, any activity is an opportunity for informal practice, and a system might remind a user to be present with a range of experiences, from taking out the trash to riding a roller coaster. Consistent with therapeutic practice, some technologies prompt users to practice mindful exercises when they detect stress (e.g., \textit{\href{https://spirehealth.com/pages/spire-for-mindfulness}{Spire}}). Slovak et al.'s research on scaffolding situated interventions for social emotional learning suggests that such situations require just the right emotional temperature--not `too cool' so that the stressor is non-existent, but also not `too hot' so as to overwhelm the learner's capacity to employ the technique \cite{slovak2016scaffolding}. \squeeze{Timing also matters; it is often better to provide interventions before a stressful event rather than during it \cite{Sano2017-bi}. Similarly, there might be opportune moments at which it is appropriately challenging--perhaps stressful, but not \textit{too} stressful--for people to practice applying mindfulness as a self-regulation technique.}

\subsection{Designing for Coherence}
In mindfulness classes, the teacher creates a container for formal practice. In mindfulness technologies, the designer assumes part of this important responsibility. 

Because teachers repeatedly called out how certain design conventions in apps--such as ads and gamification--undermined the tenets of mindfulness, we focus on the challenge of achieving \textbf{coherence} in the container for mindfulness technologies.

\squeeze{We define coherence in design as the extent to which the underlying principles of a technology and its features make sense together. This relates to Alexander's goodness of fit, wherein the purpose of design is to find a fit between form (a design solution) and context (``that part of the world that makes demands of the form'') \cite[p.19]{Alexander1964-si}. Coherence draws attention to an additional fit, that between a design solution and the underlying principles of the practice it represents (e.g., mindfulness),} that may be helpful to establish before moving on to achieve Alexandrian goodness of fit with the wider demands of the world.

\squeeze{When considering mindfulness technologies, it was easy for teachers to note cases where this coherence was violated. This too aligns with Alexander's observation that it is easy for architects to identify cases of bad fit, but that good fit is harder to find and requires "a negative process of neutralizing the incongruities, or irritants, or forces, which cause misfit" \cite[p.24]{Alexander1964-si}.}


Design metaphors are one way that designs can create--or violate--coherence. For instance, for mindfulness in its role as a mediator, a linear progress bar is apt. As a goal, mindfulness could be represented as journey along a winding path that never reaches the end. As a way of being, mindfulness could fit match with a cyclical process that has no beginning or end, such as the changing of the seasons. 




\squeeze{Before deciding upon the concept of coherence, we considered alternatives such as consistency and fidelity, however these mischaracterize what mindfulness teachers found missing in some apps. Design education emphasizes the importance of \textit{consistency} in visuals, voice, and interaction patterns \cite{Soegaard2019-hx}. However, the issues that mindfulness teachers pointed to were at a higher-level than mismatched color palettes or inconsistent feedback.} For a written text to achieve coherence, it must make sense not just from sentence-to-sentence, but as a unified whole. So too with technology that seeks to cultivate a quality like mindfulness in the user: its features must not only fit together, but also adhere to an underlying principle.



\squeeze{Another candidate concept, \textit{fidelity}, describes faithfulness to the original. Because of the long history of mindfulness in Buddhism, we expected some mindfulness teachers to be critical of unfaithful digital representations. And teachers did feel that mindfulness apps should faithfully embody the \textit{principles} of the practice, e.g., that social features should be designed for non-judgmental awareness. However, they largely held that technologies need not follow the \textit{form} of traditional mindfulness practice, i.e., that the only correct container for attaining mindful presence is sitting meditation with cushions and candles. In the same way that telecommunications researchers moved past regarding face-to-face communication as the experience that all other technologies should reproduce and recognized the unique affordances of the new digital medium \cite{Hollan1992-nh}, designers of mindfulness technologies too have an opportunity to not just mirror offline practices but to explore new forms.}




At times, a deliberate break from coherence can be called for \cite{Chen2018-qz,Gaver2003-db,Mekler_undated-md}. Artifacts might be more conducive to mindfulness if they offer a break from the tired patterns of the routine \cite{Akama2015-ey, Salehzadeh_Niksirat2017-mj}. This is also true in designing for contemplative practices as a complement to mindfulness. In constructive practice, contemplating one’s own mortality is a common way of bringing one's core values into focus \cite{Dahl2015-am}. Seeing computer-generated images of the future self increases savings behavior \cite{Hershfield2011-nn}; designers might leverage similar approaches as part of meditations that seek to cultivate virtuous traits. In deconstructive practice, Zen Koans are intended to baffle the intellect and instill a master’s state of mind in the student \cite{Zug1967-fc}. Similarly, a mobile app might introduce a leaderboard of the most practice sessions one week, only to arbitrarily flip the ordering of the leaderboard the next week, as part of a lesson that prompts students to question the importance of being at the top of the leaderboard anyway \cite{Munson2012-ez}.

For designers then, crafting a coherent--or deliberately incoherent--design requires identifying alignment between the underlying principles of mindfulness and the features of their technology.  \squeeze{To do so, designers might draw upon value sensitive design \cite{Friedman2013-ia}, using its conceptual and empirical investigations to clarify which mindfulness principles are relevant to their work. Participatory design  \cite{Spinuzzi_undated-ln} with mindfulness teachers might also be a practical way to bootstrap such an understanding.}







\section{Conclusion}
\squeezemore{Mindfulness has struck a chord with designers in industry and academia \cite{Mani2015-kv,Terzimehic2019-eo}. Grounded in our interviews with mindfulness teachers and informed by Dahl and Davidson's work \cite{Dahl2018-fc,Dahl2015-am}, we encourage technology designers to bridge from mindfulness to complementary contemplative practices that cultivate kindness and self-insight. Teachers also called for apps to scaffold the transfer of mindfulness into everyday life, so that people learn to practice outside of the tool. Finally, designers might explore mindfulness in new forms, but still strive for coherence with the underlying principles of mindfulness as a practice.}

\section{ACKNOWLEDGEMENTS}
We thank Jasmin Niess and Pawel Wozniak for discussions of prior work; Carly Maletich for feedback on the study design; Harini Gopal, Rose Guttman, Holly Hetherington, Daiana Kaplan, Julie Mills, Lexi Rohrer, Emily Rosenfield, and Anna Schmitz for reviewing apps; and all of the mindfulness teachers in the study. This research was funded in part by the National Science Foundation under award IIS-1553167.

\balance{}

\bibliographystyle{SIGCHI-Reference-Format}
\bibliography{sample}

\end{document}